\documentstyle[prl,aps]{revtex}
\begin{document}
\renewcommand{\thefigure}{\arabic{figure}}
\def\reflabel#1{\noexpand\llap{\string\string\string#1\hskip.31in}}
\draft
\flushbottom
\twocolumn
[
\title{%
\bf Avalanche Dynamics of Crack Propagation 
 and Contact Line Depinning
}

\author{%
Peter B. Thomas and Maya Paczuski }
\address{
Department of Physics, Brookhaven National Laboratory, 
Upton, New York, 11973 \\
email: thomas@cmth.phy.bnl.gov maya@cmt1.phy.bnl.gov \\
submitted to Phys. Rev. Lett.  Jan 31, 1996  \\
}

\maketitle
\tightenlines
\widetext
\advance\leftskip by 57pt
\advance\rightskip by 57pt

\begin{abstract}
A model for crack propagation and contact line depinning 
is studied. Although the model contains nonlocal interactions, it
obeys general scaling relations for depinning via
localized bursts or
avalanches.  Our numerical result for
the roughness exponent in one dimension,
$\chi=0.49 \pm 0.05$, agrees with recent
experiments on cracks measuring the in-plane roughness $\chi \simeq
0.5 - 0.6$, as well as mean field arguments giving $\chi=1/2$, but is
significantly higher than the functional renormalization group
prediction $\chi = 1/3$.

\end{abstract}
]

\narrowtext
\tightenlines

Despite its enormous practical importance and intense efforts to model
the phenomenon, the physics of crack propagation is poorly understood
\cite{background}.  Recently, Bouchaud {\it et al.}  \cite{bouchaud}
proposed that crack propagation may be related to the dynamics of
interface depinning by quenched random impurities \cite{halpin}.
Subsequently, Schmittbuhl \cite{sch} {\it et al.} argued that the
propagation of a crack in a certain idealized geometry has the same
in-plane dynamics as contact line (CL) depinning on a dirty substrate.
This mapping would apply specifically to a slowly advancing crack
confined within an easy plane between two elastic solids that are
being slowly pulled apart.  They numerically measured an in-plane
roughness index $\chi \simeq 0.35$, in seeming agreement with the
functional renormalization group (RG) \cite{rg,nf} prediction by Erta{\c s}
and Kardar \cite{ek} that $\chi=(2-d)/3$ exactly \cite{comment2}.
Daguier, Bouchaud, and Lapasset \cite{dbl} have actually
measured the in-plane roughness index of slowly moving cracks in
metallic alloys and for stopped cracks found $\chi \simeq 0.5 -
0.6$, in apparent contrast to the theoretical predictions.  Since the
model \cite{sch} clearly represents a vastly simplified description of
crack propagation, the discrepency might be due to an inherent
deficiency.  Here we present results which lead to a different
conclusion.

Our main numerical result from simulations of the model is $\chi =
0.49 \pm 0.05$ with scaling over almost three decades in length scale.
Our result is inconsistent with the functional RG prediction, but
compares reasonably well with the experimental finding.  We also show
that despite the nonlocal interactions in the model motion near the
depinning transition takes place in terms of localized bursts, or
avalanches.  As a result the critical dynamics obeys general scaling
relations for avalanche phenomena in systems out of equilibrium
\cite{pmb}.  Finally, based on numerical evidence, we conjecture that
the critical exponents for CL depinning and the related model for
crack propagation are given by a simple mean field theory giving
$\chi=1/2$, $D=3/2$, $\tau=4/3$, $\nu=2$, $\gamma=2$, $z=1$,
$d_f=1/2$, $\beta=1$, and $\pi=2$ in one dimension.

Applying a sufficient force to an interface causes it to move through
a random medium with a finite velocity which vanishes continuously at
a depinning transition where quenched impurities pin the interface
into a static configuration.  Rather than exhibiting a smooth
continuous motion, the dynamics near the transition point takes place
in terms of intermittent, localized bursts.  A scaling theory has been
developed that relates the critical dynamics of interface motion near
the depinning transition to the spatiotemporal structure of avalanches
\cite{pmb}.  This theory encompasses systems with local interactions
\cite{sneppen,leschhorn} such as the motion of magnetic domain walls
 in the presence of quenched disorder, fluid invasion in a porous media,
extinguishing flame fronts
\cite{mppre}, invasion
percolation, and flux creep \cite{ip}.  Here we propose
that it also includes contact line depinning and the related phenomena
in crack propagation.

Unlike the examples mentioned above, CL motion is governed by a
nonlocal integral equation.  This comes about because the capillary
energy associated with small deformations of the CL has the long
wavelength limit 
\begin{eqnarray}
U_{cap}&= {\gamma \Theta^2 \over 2}{\bf \int}_{2\pi\over L}^{2\pi\over a} 
{dq\over 2\pi} |q| |h(q)|^2 \nonumber \\
      & = {\gamma \Theta^2 \over 2\pi}{\bf \int}_{a<|x-x'|<L} 
{\big( h(x) -h(x') \big)^2 \over (x-x')^2} \quad ,
\end{eqnarray}
associated with increasing the entire area of the liquid-vapor interface,
rather than the usual line tension energy $q^2 |h(q)|^2$
\cite{jg}.  Here $h(x)$, the height of the CL profile at
position $x$, is assumed to be single valued.  $\Theta \rightarrow 0$
is the macroscopic contact angle that the liquid-vapor interface makes
with the substrate at the CL, $\gamma$ is the liquid-vapor interfacial
tension, $a$ is a small scale cutoff, and $L$ is the linear extent of
the one dimensional CL.  The unusual capillary enery has a significant
effect on the observed height fluctuations both in equilibrium and out
of equilibrium \cite{jg}.

If the CL is driven slowly and most of the dissipation
occurs in its vicinity, the capillary energy enters into the equation
of motion for the height profile as an applied force $F_{int}(x,t) =-
{\delta U_{cap} \over \delta h(x,t)}$, where $t$ is time.  In
addition, there is a random contribution to the local force density
from the quenched impurities $\eta\big(x,h(x,t)\big)$.
The CL is driven with an
external force $F$ \cite{ek}.

In our numerical simulations the variables $(h,x,t)$ are discretized
to be integers.  We impose a geometry where the CL is
periodic in $x$, i.e. $h(x + mL)= h(x)$, where $m$ is
integer and L is the system size studied.  This requires a summation
of the nonlocal kernel over all periods and modifies the $1/x^2$
interaction to be $\big({\pi \over L}\big)^2/\sin^2\big(\pi x / L\big)$.
In the
thermodynamic $L \rightarrow \infty$ limit the summed interaction simplifies
to $1/x^2$.  The total force density $f(x,t)$ 
\begin{equation}
 = -\nu \big({\pi\over L}\big)^2 \sum_{x' \neq x}
{ (h(x,t) - h(x',t))\over \sin^2 \big( {\pi(x-x') \over
L}\big)} - \eta(x,h) + F \quad ,
\end{equation}
where random forces $\eta$ are chosen independently from a flat
distribution between zero and one, and the sum is over all lattice
sites.  Essentially the same
equation was proposed \cite{sch} to describe the in-plane
dynamics of a crack propagating
within an easy plane of an elastic block that is being slowly wedged
open.  Of course, this represents a very idealized picture for crack
propagation since, for instance, it ignores the out-of-plane meandering of the
propagating crack and the out-of-plane
roughness of the resulting crack surface \cite{background}.

A constant force
depinning transition is implemented as follows: At each time step, the
force at each site  is evaluated.  If  $f(x,t)  >0$, then the site  is
unstable or ``active'' and the height at that site  is advanced by one
unit $h\rightarrow h+1$; otherwise  the site is  pinned and the height
is  unchanged.   After the heights  at    all active  sites have  been
advanced by one  unit, time is   advanced by one unit  $(t \rightarrow
t+1)$, the local forces are re-evaluated, and the process is repeated.
For $F > F_c$, the interface moves
with  a finite average velocity $v   = n_{act}/L$ where $n_{act}$ is
the average number of unstable, moving sites.  

The critical value $F_c$ can be found by locating the site with the
largest value of $f(x,t)$ in a pinned configuration and increasing $F$
until that site becomes unstable.  This generates a burst of activity,
or an avalanche, during which the interface moves.  Eventually the
burst dies out and the interface becomes stuck, with all sites frozen
in a new configuration.  The closer $F$ gets to $F_c$, the larger is
the average spurt of growth, or avalanche size, $s$.  The average
avalanche size diverges at the depinning transition $F =F_c$.  In
order to obtain statistics for the steady state properties, one can
set a value $F = F_c - \Delta F$, where $\Delta F \ll 1$.  At a moment
in time when there are no active sites with $f(x,t)>0$, the current
$F$ avalanche stops.  Then a new $F$ avalanche is initiated by
advancing the height of the site with the  largest value $f(x,t)$
by one step.  Starting from a flat configuration, $h(x,t=0)=0$ for all
$x$ and repeating this process many times, the CL will reach a steady
state in a time $t_{ss} \sim L^z$.  All of the steady state
properties of the depinning transition studied in this work were
obtained using this method.

In Fig. 1, we show an actual realization of the height profile in the
steady state near the depinning transition. 
The method  of estimating the roughness exponent
$\chi$ which works   best for this model,
is to calculate the power spectrum $P(k)$ \cite{halpin,sch}
for the  height profile
in the steady state. 
Fig. 2 shows
power law behavior $P(k) \sim k^{-1-2\chi}$.  The asymptotic slope is
$-1.98$, which fits the data over 2.5 decades and gives the result
$\chi = 0.49$.

One can compare Fig.  2 to the equivalent Fig. 3. in Ref.
\cite{sch} where a slope for $P(k)$ of $-1.7$ was measured with
much fewer samples
on a somewhat smaller system size $L=2048$ than 
we study.  On inspection of their figure, it is apparent that there is
a systematic deviation from their fitted slope at small values of $k$
(large values of $x$) where the asymptotic behavior dominates.  Their
data points at small $k$ do not fall near the ``best fit'' line and
actually give an apparent slope which is closer to what we measure.

The equal time, height-height correlation function is
\begin{equation}
C(r) = < (h(x+r,t)- {\overline h(t)})(h(x,t) -{\overline h(t)}) >
\quad ,
\end{equation}
where the   angular brackets denote   an average  over  time $t$ and
over $x$, and $\overline h(t)$ is the average height of the interface
at time $t$.
  In
Fig. 3, we plot  $C(r)/L^{2\chi}$ versus $r/L$, where $\chi$
assumes its mean field value
1/2, so that $C(0) \sim L$.
  The large
deviation for small system size $L=128$ for $r$ near zero indicates
that such a small system size is not in the asymptotic regime.  However
we find reasonable good data collapse for $L=1024 - 8192$.
Our main numerical result is that $\chi$ is bounded by $\chi = 0.49
\pm 0.05$.

We now discuss the scaling behavior of avalanches  in the
model.  The size   $s$  of an  $F$
avalanche is   the integrated motion,  or the area
between the initial configuration  and the final configuration
of the avalanche.  In an
experimental situation,  perhaps, $F$ would  be increased slightly for
each subsequent  avalanche, but for a  sufficiently large system, many
avalanches would occur within a narrow interval  centered on $F$.  The
statistics of those   avalanches   are described by   the  probability
distribution $P(s,\Delta F)$.

In analogy with other depinning  problems \cite{paralleltl,mppre,pmb}, it
is plausible that the  probability distribution of avalanche sizes is
given by
\begin{equation}
P(s,\Delta F) \sim s^{-\tau}g(s\Delta F^{\nu_s}) \quad
\end{equation}
near the  critical point $\Delta  F =0$.  Fig.   4  shows the measured
distribution for small  $\Delta F$; it decays as a
power law  over four decades
with a characteristic exponent $\tau = 1.31\pm 0.06$ up to a
cutoff  determined  by the system   size.

In  spite of the  non-local nature  of  the interactions, we find that
similar   to other    interface   models  which  have   been   studied
\cite{sneppen,mppre,pmb}, {\it the  dynamics during an avalanche
is local}.   Fig. 5 shows  the  location of the  active, moving
sites vs. time, for one  avalanche of moderate size.  It is clear
from the figure that the avalanche spreads out in time.  In fact,
the distance spread $r(t)$ grows with time as $r(t)\sim t^{1/z}$.  If the 
projection of the completed avalanche onto the sites that
it covered is compact, and the avalanche size
$s \sim r^D$, where $r$ is the spatial
extent of the avalanche, then
$D = d+\chi$ \cite{pmb}.   In general $D\neq z$.
For the same avalanches used in Fig. 4,
we  measured $D=1.5\pm 0.15$ which is
consistent with the scaling relation and our result
for $\chi$.

We now discuss a simple mean field theory to explain these results.
The characteristic dimension of the nonlocal interaction term in
Eq. (2) at length scale $l$ is $h/l$ where $h \sim l^{\chi}$.  The
sum of random forces at this scale makes a contribution $l^{1/2}/l$
per unit length.  Balancing these two terms gives $\chi=1/2$.
Since the velocity $v(l) \sim h/t \sim f(l)$, the time $t \sim l$
so that the dynamical exponent $z=1$.  Defining the exponent
$\beta$ by $v \sim \Delta F^{\beta}$ and the cutoff $l_{co} \sim
\Delta F^{-\nu}$, one obtains the  scaling relation
$\beta=\nu(z-\chi)$ \cite{nf}.  The exponent $\beta$ can be obtained by
integrating the equation of motion over the entire system, so that
the nonlocal part vanishes.  This gives
\begin{equation}
v=F + {1\over L}\int dx \eta(x,h)= F + \overline \eta(F) \quad ,
\label{velocity}
\end{equation}
where the average $\overline \eta(F)$ over the CL is evaluated at
a particular value of $F$ in the steady state.
At $F=F_c$, $v=0$ so that $\overline \eta(F_c) = -F_c$.
In general for $F=F_c + \Delta F$ we have 
$\overline \eta(F) = -F_c + g(\Delta F)$.
Substituting this last result into Eq. (\ref{velocity}), we get
\begin{equation}
v= F - F_c + g(F - F_c) \quad .
\end{equation}
Assuming $g$ has a power series expansion gives $\beta =1$ 
The scaling relation for $\beta$ then gives $\nu=2$\cite{comment1}.  

In general, for systems with local avalanche dynamics the exponent $\tau$ for
the distribution of avalanche sizes obeys the relation $\tau = 1 + {d
- 1/\nu \over (d+\chi)}$ \cite{pmb}, and substituting the mean field
values for one dimension $(d=1)$ gives $\tau=4/3$ in agreement with
our numerics.  The exponent $\gamma$ for the divergence of the average
avalanche size is $\gamma=\nu D (2-\tau)$, and $\gamma=2$ in mean
field.  In the constant velocity depinning transition the critical
exponent $b$ in Ref. \cite{sch} characterizing the power law
distribution between subsequent activity obeys $b=\pi-1=D(2-\tau)$
according to the scaling theory \cite{mppre,pmb}.  The exponent
$b$ was measured to be $b=0.9 \pm 0.05$ \cite{sch} which is consistent
with the scaling relation and our numerical values for $D$ and $\tau$.
The mean field prediction is that $b=\pi-1= 1$.  Also, in the constant
velocity case, there are scaling relations for the fractal dimension
of active sites $d_f=D(\tau -1)$ and the dynamical exponent
$z=D(2-\tau)$ \cite{pmb} which in mean field take the values
$d_f=1/2$ and $z=1$.  This last result suggests that {\it all} of the
critical exponents are the same in the constant velocity and constant
force case for this problem.

In addition to CL depinning and crack propagation, exact
results have been put forward using the functional RG for interface
depinning with purely local interactions, where Narayan and Fisher
predicted $\chi = (4-d)/3$ \cite{nf}.  Numerical simulations of
related lattice models in both one and two dimensions give $\chi
=1.23$ \cite{pmb} ($\chi=1.25$ \cite{leschhorn}) in $d=1$ and $\chi =
0.72$ in $d=2$ \cite{pmb}.  In all cases thus far where a comparison
can be made, including our result,
the numerical simulations give a roughness exponent
higher than the functional RG prediction. 

 Our theoretical picture for depinning phenomena is based on avalanche
dynamics.  To our knowledge, avalanches are not explicitly contained
in the functional RG theory.  We suspect that the apparent failure of
the functional RG to reproduce numerical results for
interface depinning may express an inadequacy of these methods to
describe avalanche dynamics.  The agreement between our numerical results
and experiments measuring the in-plane roughness index of cracks
supports the model for crack propagation proposed by
Schmittbuhl, Roux, Vilotte, and M{\aa}l{\o}y \cite{sch}.

We thank V. Emery, P. Bak, and D. Dhar for useful discussions.
This work was supported by the
U.S. Department of Energy 
Division of Materials Science,  under contract DE-AC02-76CH00016. MP
thanks the U.S. Department of Energy Distinguished Postdoctoral Research
Program for partial financial support.


\def\fig#1\par{\begin{figure}\caption{#1}\end{figure}}

\fig\label{one}%
An actual realization of the height profile  $h(x,t)$
for $L=8192$ [20]. 

\fig\label{two}%
A log-log plot of the power spectrum of the height profile versus $k$
for $L=8192$.  The height profile was sampled every
$t=273$ time steps and over $10^6$ samples were used to compute $P(k)$.
The straight line is a best
fit to all the points $k<0.1$. It has a slope of $-1.98$.

\fig\label{three}
Rescaled plot of the equal time height-height  correlation function
$(C(r)/L)$ vs. $r/L$
for  $L=128$, $F=0.83$  (open circles), $L=1024$,  $F=0.828$
(triangles),  and $L=8192$ (filled circles).
This collapse of the data for $L=1024-8192$
is consistent with $\chi =1/2$.

\fig\label{four}
The avalanche   size distribution near   the depinning  transition for
$L=8192$.  Over $5\times 10^4$ avalanches were measured.
The slope gives an estimate of
$\tau = 1.31 \pm 0.06$. 

\fig\label{five}
The location  (marked  as  dots) of  the  active, moving  sites   as a
function of time,  for $L=1024$,  during an
an avalanche which begins at $t=0$ and
ends at  $t=545$. The entire avalanche has a size $s=9814$.
The avalanches  are local and spread out in time as $r\sim t^{1/z}$.


\begin{thebibliography}{99}
\bibitem{comment1} Some of these  mean field values 
were previously found in Ref. \cite{ek}.
\bibitem{comment3} For all simulations of Eq. (2) for system
size $L=8192$, we used 
$\nu=0.01$ and
$F=0.8294$.
\end{thebibliography}
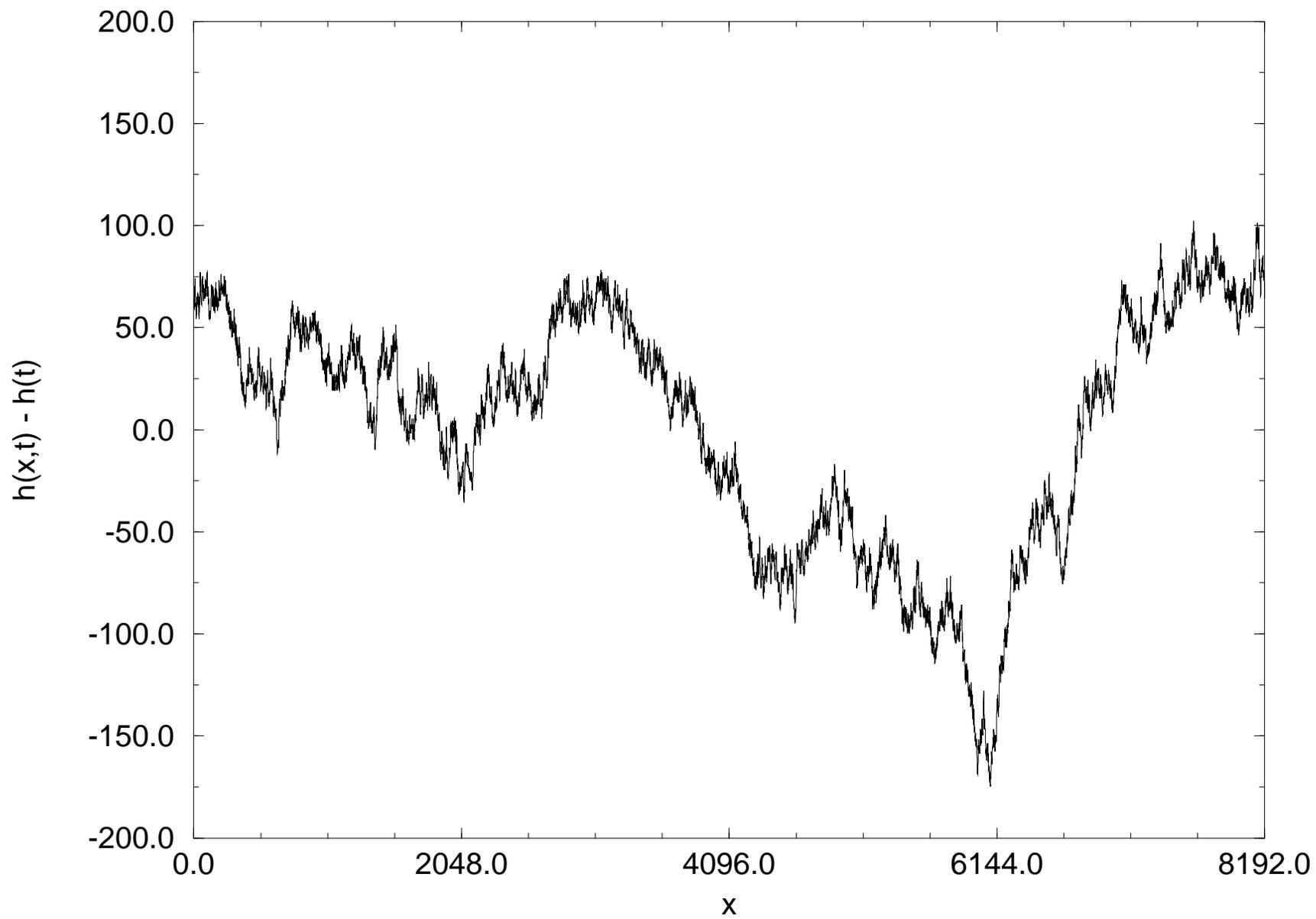
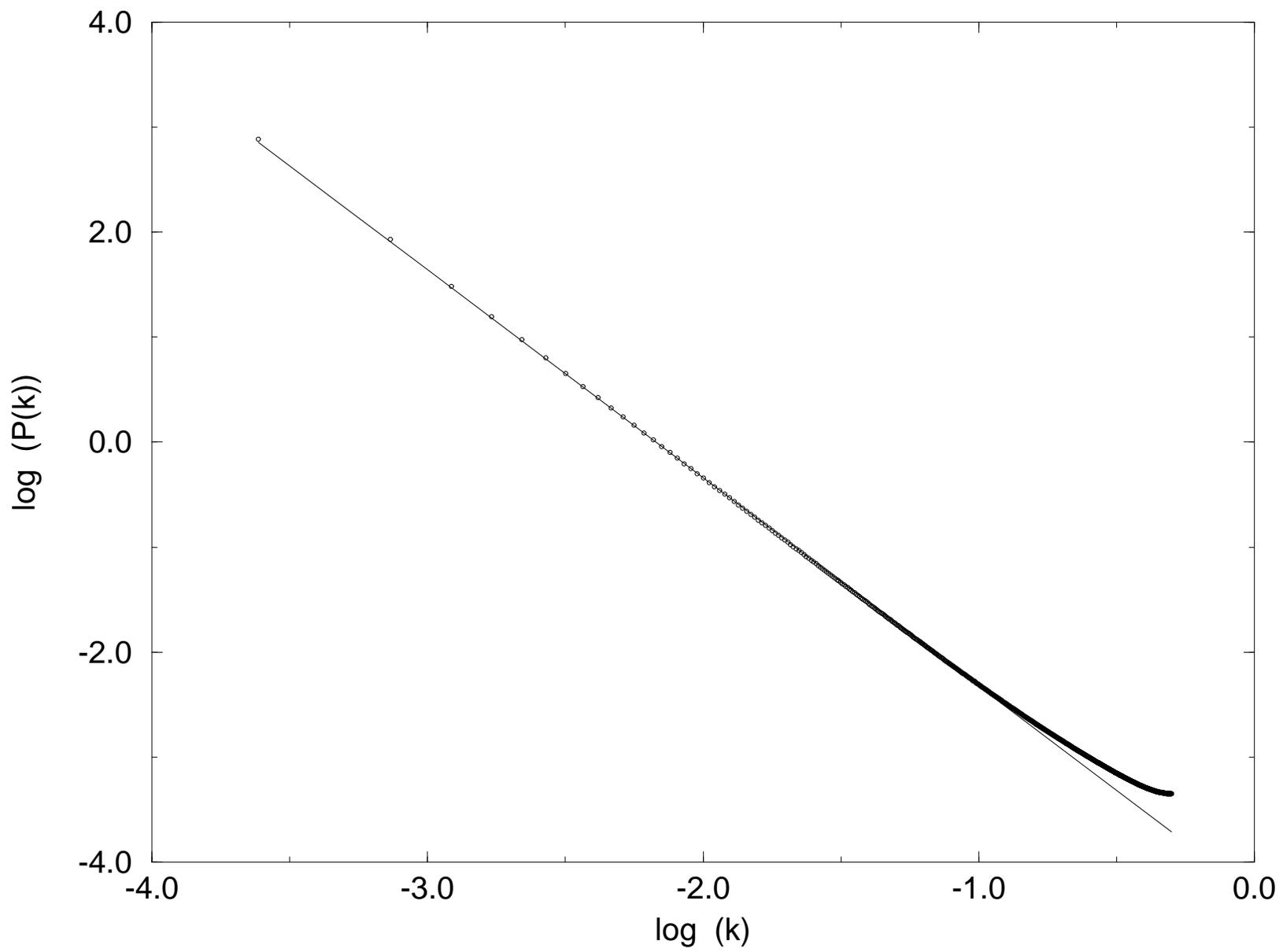
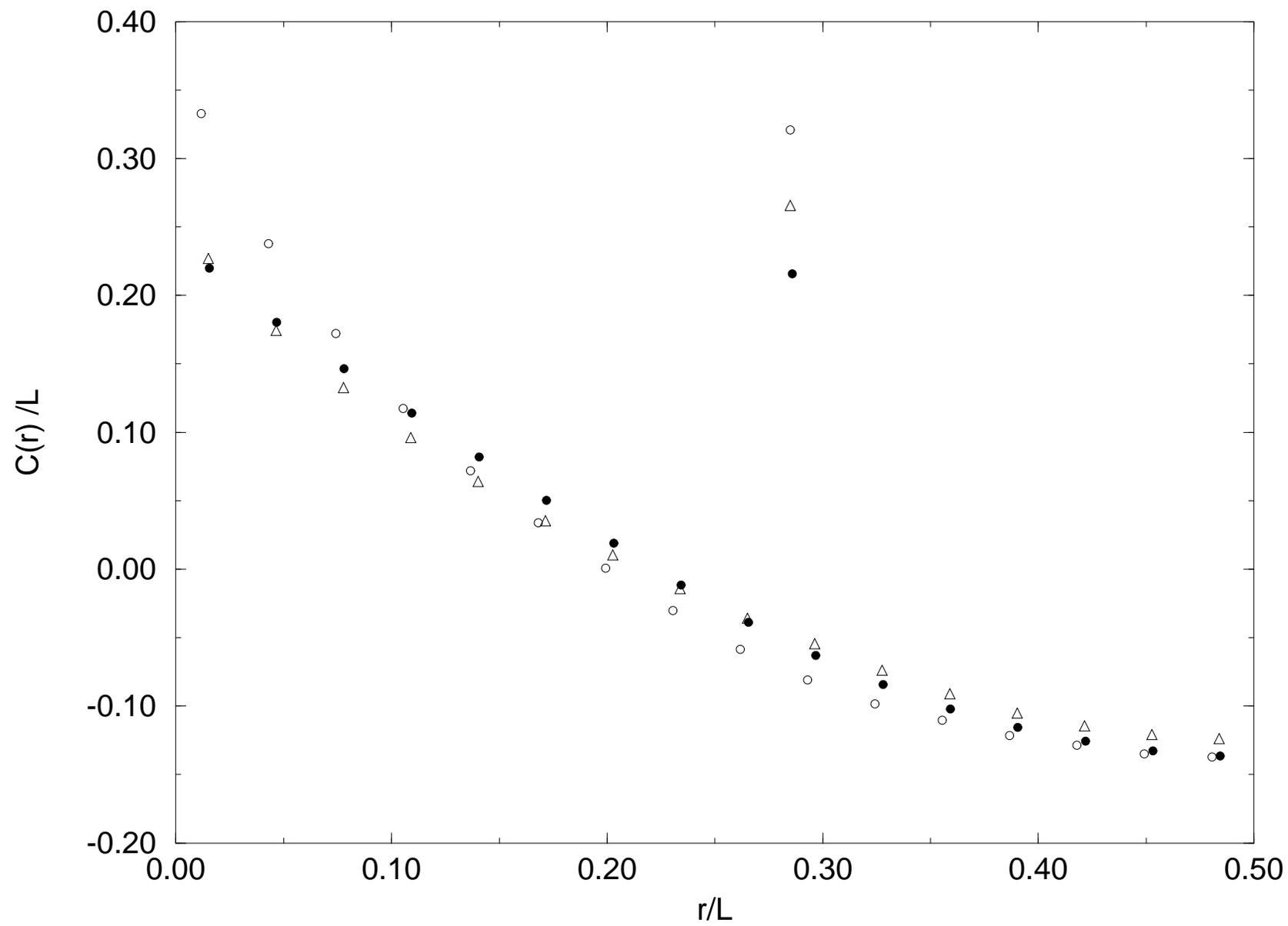
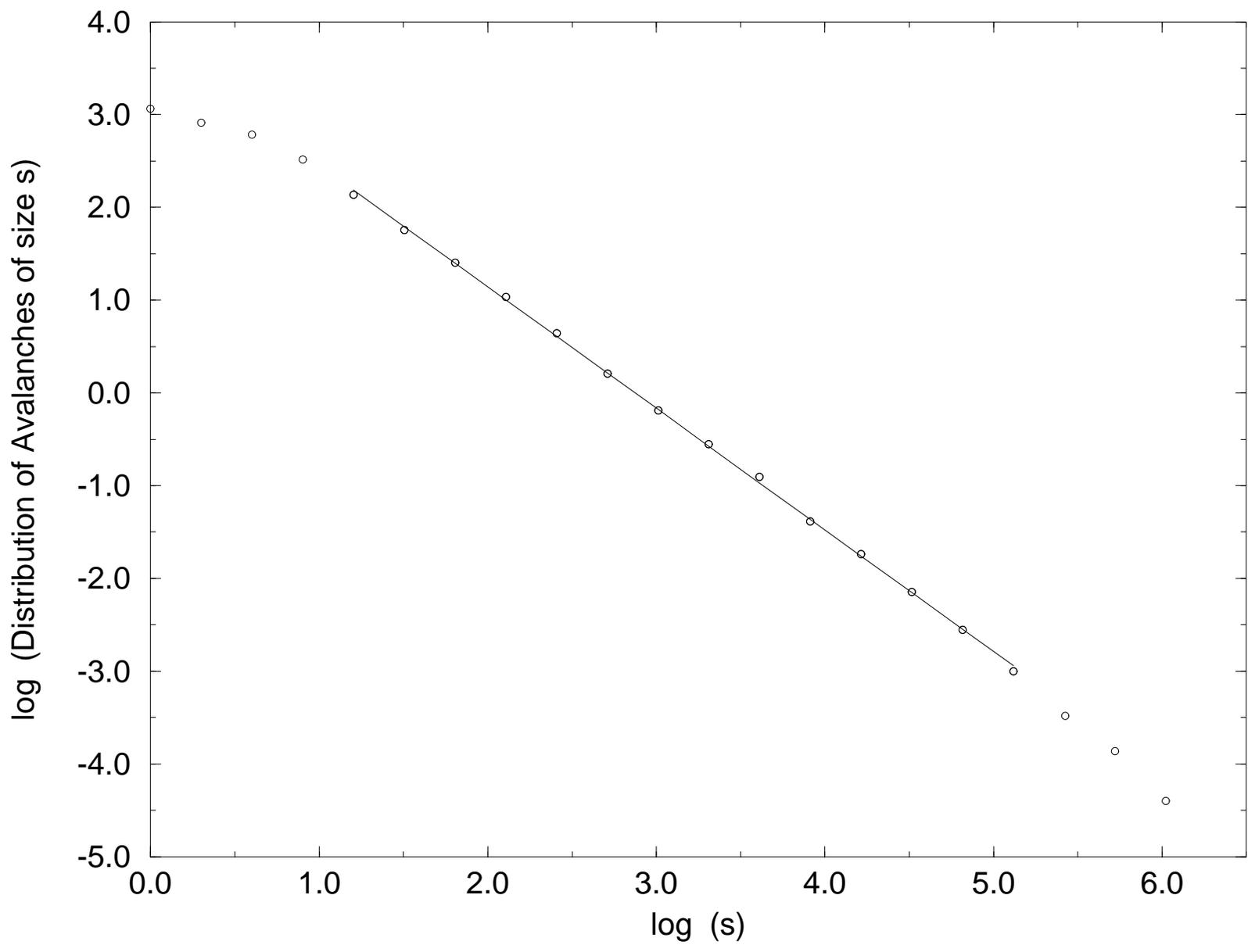
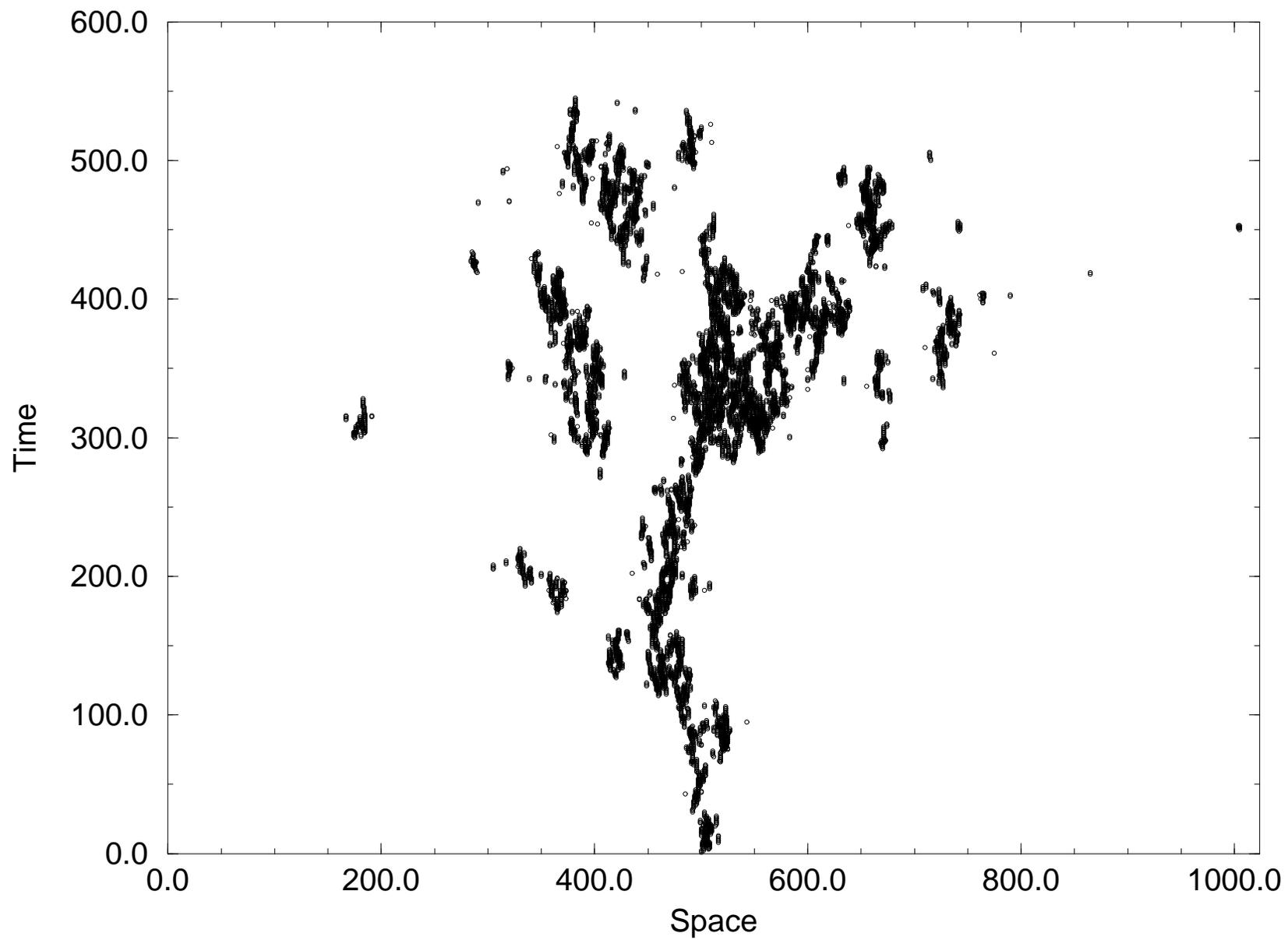
\end{document}